\documentclass[preprint,12pt]{elsarticle}

\usepackage{amssymb}

\usepackage{epstopdf}

\journal{Journal of Alloys and compounds}

\begin{document}

\begin{frontmatter}



\title{Nematicity, magnetic fluctuation and ferro-spin-orbital ordering in BaFe$_2$As$_2$ family}

 \author[1]{Smritijit Sen}
\author[1,2]{Haranath Ghosh\fnref{email}}
\address[1]{Homi Bhabha National Institute, Anushaktinagar, Mumbai 400 094, India.}
\address[2]{Indus Synchrotrons Utilization Division, Raja Ramanna Centre for Advanced Technology,
Indore -452013, India.}
\fntext[email]{Corresponding author : hng@rrcat.gov.in}


\begin{abstract}
Through detailed electronic structure simulations we show that the electronic orbital ordering
(between d$_{yz}$ and d$_{xz}$ bands) takes place due to local breaking of in-plane symmetry
that generates two non-equivalent $a$, $b$ directions in 122 family of Fe-based superconductors.
Orbital ordering is strongly anisotropic
and the temperature dependence of the corner zone orbital order maps to that of the
orthorhombicity parameter. Orbital anisotropy results in two distinct spin density wave nesting
wave vectors and causes inter-orbital charge and spin fluctuations. Temperature dependence of the
orbital order is proportional to the  nematic order and it sets in at a temperature where
magnetic fluctuation starts building. Magnetic fluctuations
 in the orthorhombic phase is characterized through evolution of Stoner factor
 which reproduces experimental findings very accurately.
Orbital ordering becomes strongly spin dependent in presence of
magnetic interaction.  Occupation probabilities of all the Fe-d-orbitals exhibit temperature
dependence indicating their possible contribution in orbital fluctuation. This need to be
contrasted with the usual definition of nematic order parameter (n$_{d_{xz}}$-n$_{d_{yz}}$).
Relationship among orbital fluctuations, magnetic fluctuations and nematicity are established.
\end{abstract}


\end{frontmatter}


\section{Introduction}
\label{Introduction}

The discovery of high temperature superconductivity in Fe-based materials attaining T$_c$ 
as large as 109K \cite{Ge}, has lead to a huge up surge of research 
for further discovery of such new materials \cite{Hosono}.  Seven years after its discovery, 
while a clear consensus on the mechanism of superconductivity has not yet  been reached,
 understanding on the structural, 
magnetic transitions and their mutual influences on superconductivity remain central issue of frontier research
\cite{Hosono,Fernandes,Wu}. A large number of undoped Fe-based materials show spin density wave (SDW) magnetic state
whose transition temperature coincides with that of the structural transition 
(which gets separated through doping as well as 
pressure). Both the transitions being second order 
in nature, can have a conflict with Landau theory of phase transition unless there would be a precursor transition
at higher temperatures. According to Landau theory, occurrence of two simultaneous transitions may be purely coincidental,
mutually independent, or one of the transitions be first order type or there must be a precursor to one of the transitions at a higher temperature. What is that precursor ?

Origin of the structural transition is not purely lattice driven but an
electronic one; the orbital ordering of Fe d$_{yz}$ and d$_{xz}$ orbitals \cite{Yi,sust} is the
primary reason for structural transition.
Among some of the normal state properties, transport in "preferred" direction has been 
observed unambiguously in many experiments \--- inelastic neutron scattering (INS) \cite{Zhao}, scanning tunnelling 
microscope, impurity \cite{Allan}, resistivity \cite{Chu}, optical conductivity \cite{Dusza}, angle
resolved photo electron spectroscopy \cite{Yi} and so on.
Overall, origin of such phenomena is related to the breaking of four-fold rotational
(C$_4$) symmetry of the tetragonal phase known as nematicity --- the precursor. 
Whether the origin of nematic phase is spin driven or orbital driven 
is far from being settled.
The nematic phase is observed in FeSe materials, that has structural transition at 90 K 
but no trace of long-range magnetic order \cite{Song} 
indicating nematicity is orbital fluctuation driven \cite{Baek}.  However, observation
of an additional C$_4$ phase deep inside the orthorhombic (C$_2$) phase in Ba$_{1-x}$Na$_x$Fe$_2$As$_2$ 
close to the suppression of magnetic spin density wave (SDW) order favours magnetically driven
nematic order \cite{Avci}.
Role of nematic phase as regards to the
mechanism of superconductivity or symmetry of Cooper pair wave function is not straight forward \cite{shimo} 
but the fact that spin fluctuation 
leads to s$^{+-}$ superconductivity \cite{Mazin} whereas the orbital fluctuation leads to s$^{++}$ superconductivity in Fe-pnictides \cite{Kontani}
are established and magnetism competes with superconductivity \cite{Ghoshepl,RMF1}.
 There are also evidences of 'nematic order' in the pseudogap phase of the other class of high temperature cuprates superconductors which are known to be strongly correlated materials \cite{Lawler,Hinkov,Daou,Fujita}.
 On the other hand, 122 family of Fe-based superconductors are generally considered as weakly correlated systems. Therefore, the 
 study of
 nematicity in Fe-based superconductors is of fundamental importance.
Possible origins of nematic phase are well described in \cite{Fernandes} as (a) structural 
distortion, (b) charge/orbital order, (c) spin order.
Whatever be the nematic order parameter it must couple linearly to the orthorhombic distortion 
\cite{prl}.
Nematicity introduces electronic anisotropy leading to two different nesting vectors which 
in turn leads to two competing spin density wave (SDW) instabilities (Z$_2 ~\times$ O(3) symmetry breaking) \cite{RMF2}.
Coupling of the orbital order to SDW and vice versa has been used as inputs in Ginzburg-Landau formalism which 
provides qualitative  understanding of nematic phase.
A clear first principles understanding on whether there is any direct coupling between the
magnetic (SDW) and orbital order in Fe-pnictides is absent till to date --- is the main aim of
this work. 

We show through electronic structure 
calculation that the electronic orbital ordering
locally breaks the in-plane symmetry and generate two non-equivalent $a$, $b$ directions.
In particular, we show that below structural transition (orthorhombic phase) there is a strong
orbital anisotropy along the $\Gamma-X$ and $\Gamma-Y$ polarizations; the band at X is dominantly Fe-d$_{yz}$
derived whereas that at Y, Fe-d$_{xz}$ derived respectively. This feature reproduces
correctly experimental angle
resolved photo electron spectroscopy (ARPES) observation \cite{Yi}. This is the root cause of orbital ordering
in BaFe$_{2-x}$Ru$_x$As$_2$ --- we show that the temperature dependence of the orbital ordering 
at X(Y) point reproduces exactly that of orthorhombicity parameter (hence structural transition). 
Thus structural transition is primarily electronic in origin and phonons can not be a primary order parameter 
for nematicity. Whereas the temperature
dependence of the same at $\Gamma$ point is very weak (nearly independent of temperature)
as observed experimentally \cite{Zhang}. Interestingly, Zhang et al., \cite{Zhang} observed
that the orbital splitting (ordering) at $\Gamma$ point (in case of FeSe) persists beyond 
structural transition temperature and argued that as against ferro-orbital ordering.
In order to have a complementary first principles understanding over the experimental and other studies 
\cite{Fernandes,Yi,Zhao,Baek,Avci,Zhang} we introduce magnetic interaction through tuning integrated spin density 
defined as, I$_s$=$\int$(n$_\uparrow$ (r) -n$_\downarrow$ (r))d$^3$r. In presence of finite integrated spin density I$_s$,  
spin selective orbital ordering are observed. Due to electronic orbital anisotropy (see Fig. \ref{BS}) the SDW state may be viewed as 
a superposition of two SDW states. This is because of two reasons, (i) nesting wave vector that connects nested parts of Fermi arcs along
the $\Gamma - X$ and $\Gamma - Y$ directions are different; (ii) overlap of Fe-d$_{xy}$ with d$_{xz}$ and  d$_{yz}$ is different
(below structural transition) causing anisotropic charge and spin density fluctuations. Tuning I$_s$ causes further perturbation to the
underlying SDW as well as splits the spin degeneracy of energy bands. This magnetic interaction couple with orbital (charge) fluctuations
causing further orbital anisotropy. Remarkably, this later effect is observable only in the orthorhombic phase and 
{\em not} in the tetragonal 
phase. This would be experimentally verifiable by ARPES in presence of weak Zeeman field. In presence of I$_s$ we evaluate thermal 
variations of orbital ordering at different high symmetry points. We show that the magnetic interaction
couples to the zone centre orbital ordering very strongly where as it has a substantial effect on the corner zone orbital fluctuation.  These observations support the claim by Zhang \cite{Zhang}, Fernandes \cite{Fernandes} of magnetic origin of nematic phase. Finally, we show that the orbital occupancies of all the five d-orbitals
of Fe show temperature dependencies below structural transition. This indicates to the fact that nematic order parameter may not simply
be defined as (n$_{d_{xz}}$-n$_{d_{yz}}$) but charge fluctuations from other orbitals also need to be considered, for 
example, the thermal variation of (n$_{d_{xz}}$-n$_{d_{yz}}$)+(n$_{d_{{x^2}-{y^2}}}$-n$_{d_{xy}}$) also follows orthorhombicity (see inset Fig. \ref{OOSC}c). This will put constraints on many theoretical and experimental works so far.

\section{Method}
\label{Method}
\begin{figure}[ht]
     \centering
     \includegraphics [height=08.75cm,width=8.0cm]{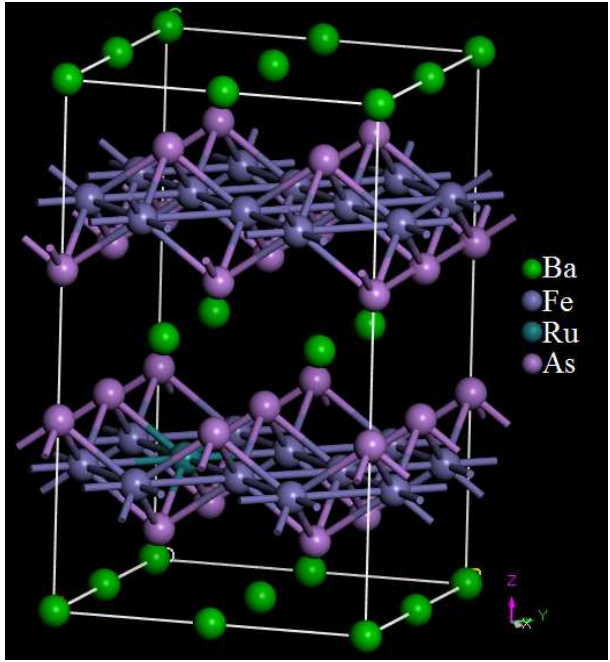}
     \caption{Structure of a 40 atoms super-cell of BaFe$_2$As$_2$, which contains 16 Fe atoms and one Ru atom corresponding to about 6$\%$ doping. Different 
     colours are used to indicate different atoms.}
     \label{SC}
     \end{figure}

First principles density functional theories can produce exact solutions of the many electron Schr\"odinger equation
 if exact electronic density is being used as input. Various modern X-ray diffraction techniques {\it e.g}., Synchrotrons radiation source etc. that determines crystallographic information at different external perturbations are essentially result of diffraction from various atomic charge densities (Bragg's diffraction). Considering experimentally determined structural parameters at different temperatures as input thus in turn provides temperature dependent densities in our first principles calculation. These input structural parameters are kept fixed through out the calculation for a fixed temperature. This is how we use a T=0 DFT formalism to bring out temperature dependent observable with the help of experimental input (energy being functional of electron density E $\equiv$ E[$\rho(r,T)]\equiv E[\rho \{a(T), b(T),c(T)\}$]. The main effect on the electronic structure from finite temperature is the underlying crystal structure, and the average crystal structure at finite temperature can usually be reliably determined from the diffraction experiment at a given temperature. This method is somewhat superior to other similar methodology, like molecular dynamics simulation (MD). Through MD simulation one finds temperature dependent lattice parameters and then use standard T=0 DFT method using GGA exchange potential (see for example, \cite{Backes,Dhaka,walsh,acta,sust} ). However, experimentally determined temperature dependent lattice parameters can be obtained with an accuracy better than 0.001 A which may not be possible in MD. Using temperature and doping dependent experimental lattice parameters $a$(T,$x$), $b$(T,$x$), 
$c$(T,$x$) and $z_{As}$(T,$x$) \cite{acta}, we obtain electronic structure
 as a function of temperature as well as 
doping, to explain the experimentally observed anomalies microscopically. One 
of the shortcomings of the density functional theory (DFT)
under generalized gradient approximation (GGA) for calculating electronic structures 
of Fe-based SCs is that it fails to reproduce accurate experimental z$_{As}$ \cite{DJSingh,dft,acta,pla}. 
This conflict arises due to strong magnetic fluctuation associated with Fe based compounds 
\cite{mazin}. This insist us to take experimental 
z$_{As}$ instead of relaxed z$_{As}$ obtained by total energy minimization, 
as one of our input parameters.
 We simulate electronic structures for both the phases, low temperature orthorhombic 
 phase with anti-ferromagnetic (AFM) as well as spin density wave (SDW) ordering and high temperature paramagnetic tetragonal phase. 
 In low temperature orthorhombic phase, along with non magnetic structures various spin configurations 
 have been employed among which the lowest energy configuration 
 is considered for electronic structure calculation.
Our first principles electronic structure calculations are carried out implementing  
 plane-wave pseudopotential method within the framework of density functional theory \cite{CASTEP}. 
In all of our temperature and doping dependent calculations the electronic exchange correlation
energy is treated under the generalized gradient approximation (GGA) using 
Perdew-Burke-Enzerhof (PBE) functional \cite{PBE}. Tackling small fraction 
of Ru substitution in place of Fe is accomplished 
by considering both, the virtual crystal approximation (VCA) 
 as well as super-cell method for convenience.
 Super-cell method is a computationally expensive method adopted to mimic finite percentage of 
 doping at a particular site. Let’s say e.g, for 5\% doping at the Fe site one needs to build a super-cell (bigger unit cell) 
 that contains 20 Fe atoms; then 1 of the Fe atoms are replaced by Ru atom (say). In the present case however, 
 a super-cell containing 16 Fe atoms (total 40 atoms) are taken out of which one is replaced by a Ru 
 (shown in FIG.\ref{SC}). This corresponds to 6\% Ru doping which is close to the experimental situation. 
 Note the size of the unit cell in the given symmetry is such that it does not allow exactly a super-cell with 20 Fe atoms. 
 Spin polarized single point energy calculations are performed using AFM 
 and SDW configuration \cite{Dai} (see inset Fig.\ref{OOSC}) for the low temperature orthorhombic phase with space group
 symmetry Fmmm (No.69) using ultrasoft pseudopotentials.
 Plane wave basis set with 
  energy cut off 500 eV and self consistent field (SCF) 
  tolerance $10^{-6}$ eV/atom has been opted for all calculations. Brillouin zone is 
  sampled in the k space within Monkhorst-Pack scheme and grid size for SCF calculation is $12\times
  12\times12$ for electronic density of state calculation in primitive cell 
  for orthorhombic phase. Band structure calculations 
  are performed along various k-paths (X, $\Gamma$ and Y) with k point separation 10$^{-3}$ {\AA}. 
Standard rotationally invariant approach due to Matteo Cococcioni and 
Stefano de Gironcoli \cite{Cococcioni} and V. I. Anisimov \cite{Anisimov} 
is used to treat the Hubbard on site repulsion effect by post-DFT 
LSDA+U method which calculates and uses the total spin-projected 
occupation of the localized manifold, as this is essential to treat the Hubbard term. 
Therefore, the method of calculation of occupation probabilities of U effected orbitals remain 
same as that of the reference given above which is also valid even in the limit U tending to zero.

\section{Results and discussions}

\begin{figure}[ht]
    \centering
    \includegraphics [height=08.75cm,width=8.0cm]{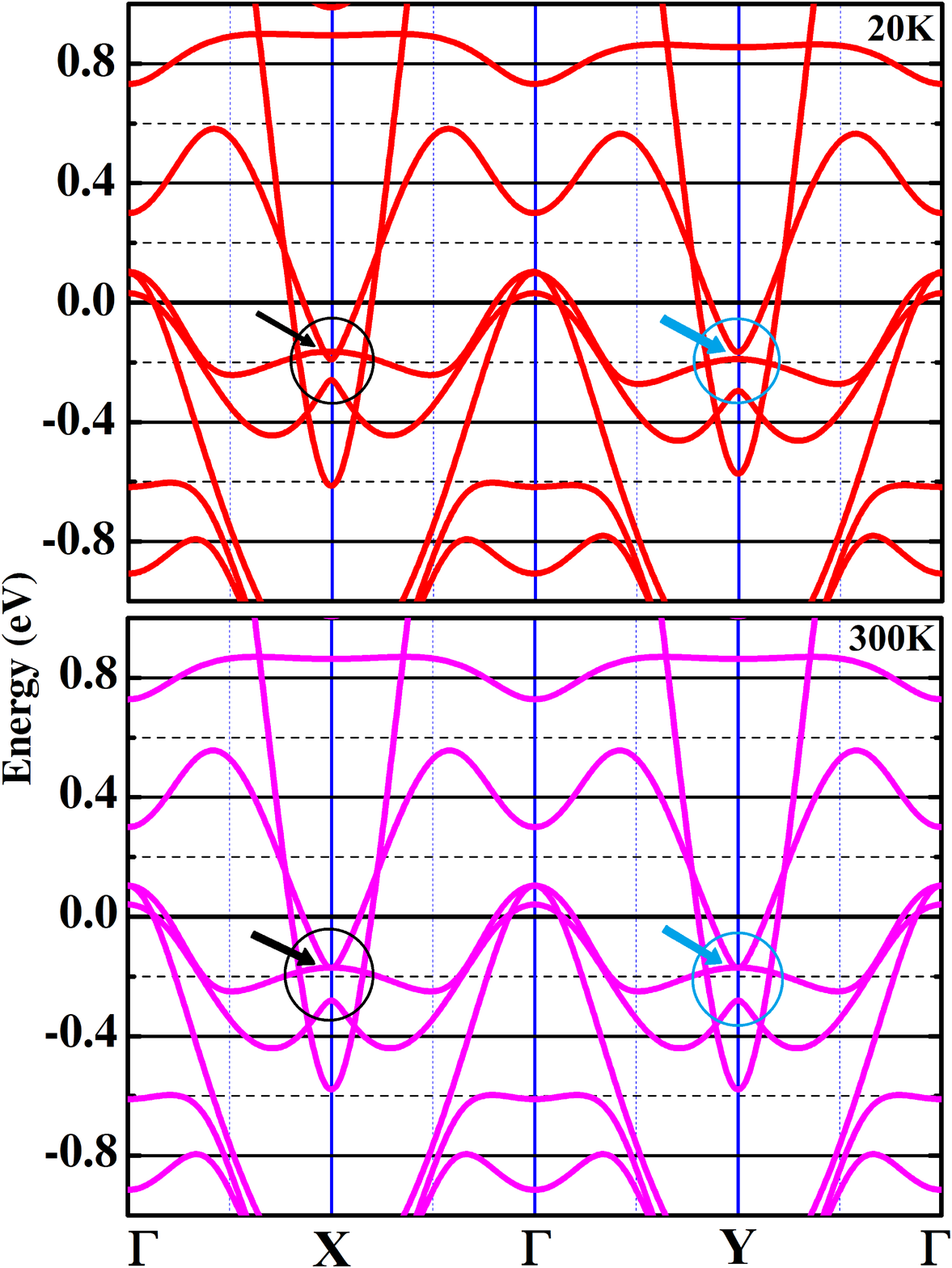}
    \caption{Calculated band structure of BaFe$_2$As$_2$ along $\Gamma-X-\Gamma-Y-\Gamma$ direction
    at 20K (red) in orthorhombic phase and 300K (magenta) in tetragonal phase. Orbital anisotropy along $\Gamma-X$ and $\Gamma-Y$ direction in the orthorhombic phase is worth noticing.}
    \label{BSF}
    \end{figure}

First, we have calculated band structures of BaFe$_2$As$_2$ system for anti-ferromagnetic (AFM) 
spin configuration (total spin zero) using experimental lattice parameters at 20K as well as 
300K along some specified k-path to probe orbital anisotropy.
Our calculated band structures of BaFe$_2$As$_2$ along the k path $\Gamma-X-\Gamma-Y-\Gamma$ at two different temperatures
corresponding to orthorhombic and tetragonal (20K and 300K) phases respectively are shown in Fig. \ref{BSF}.
Circular envelopes are drawn around X,Y
\begin{figure}[ht]
  \centering
  \includegraphics [height=6.75cm,width=8.0cm]{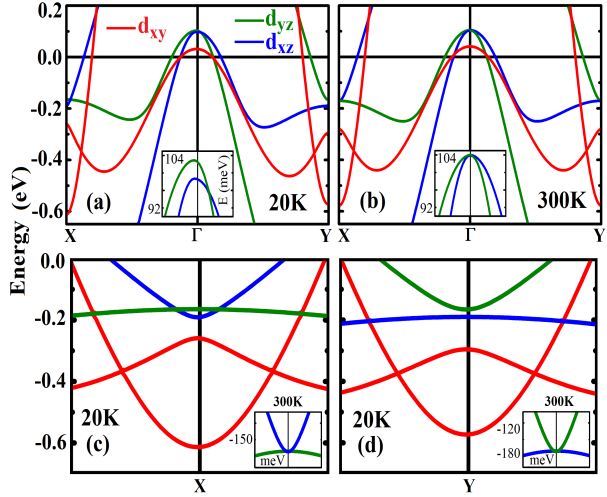}
  \caption{Calculated band structure of BaFe$_2$As$_2$ along $\Gamma-X-\Gamma-Y-\Gamma$ direction
  at 20K (upper) and 300K (lower) indicating various d orbital (d$_{yz}$, d$_{xz}$ and d$_{xy}$) using different colours. 
  Orbital ordering in the orthorhombic phase is shown in the inset figure. Electronic orbital anisotropy at the X and Y points in the orthorhombic phase is the root cause of orbital order leading to structural transition.}
 \label{BS}
 \end{figure}
 points which are then shown in Fig.\ref{BS} where splitting of $d_{xz}$/$d_{yz}$ at $\Gamma$ point has also been highlighted. 
It is very clear from Fig.\ref{BS}(a) that at 20K (orthorhombic phase), band dispersion along 
 $\Gamma-X$ direction is quite different compared to that in the $\Gamma-Y$ direction. 
In Fig. \ref{BS} the splitting of $d_{xz}$/$d_{yz}$ at $\Gamma$ point has also been highlighted 
in the inset. The same for the room temperature is then compared with. Fig.\ref{BS} demonstrate 
that the orbital ordering locally breaks the in plane symmetry and generates two non-equivalent 
$a, ~b$ directions $\perp$ to $c$. This results in two different nesting 
 wave vectors along Q$_x$ = ($\pi, 0)$ and Q$_y$ = ($0, \pi$) directions \-- that is spins are 
 parallel to each other along X-direction and anti-parallel along Y-direction (O$_3$).
 We would also like to mention that at lower temperatures there exists two Fe-Fe distances (Z$_2$) \cite{acta,sust}
 and this makes the system anisotropic both magnetically as well as in terms of band motion.
 This situation resembles to that of the nematic phase where the bilinear combination of the 
 order parameter (O$_3\times $Z$_2$) breaks the tetragonal symmetry and is invariant under symmetry transformation.
 Because of the anisotropy of the bands along X and Y directions, in general, (overlap of the d$_{xy}$ band with d$_{xz}$ and
 d$_{yz}$ bands are specially different) causes inter-band charge and spin fluctuations, which may cause for example,
  coupling between them resulting in different amplitudes of the SDW along Q$_x$ and Q$_y$ directions. Energy orderings of
 the non-degenerate $d_{xz}$/$d_{yz}$ bands sets in orbital ordering \cite{sust,shimojima,Suzuki},
 \begin{figure}
  \centering
  \includegraphics [height=5.00cm,width=6.50cm]{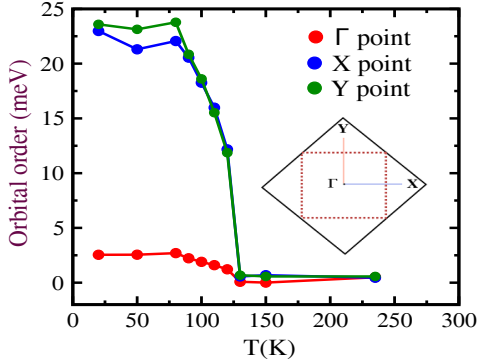}
  \caption{Calculated orbital order (in meV) around X (blue), Y (green) and $\Gamma$ (red)
   points as a function of temperature for 5$\%$ Ru doped BaFe$_2$As$_2$. Brillouin zone of orthorhombic
   BaFe$_2$As$_2$ has been shown in the inset of the figure indicating various k points (X, Y and $\Gamma$).
   The temperature dependence of orbital order is same as that of orthorhombicity ($\delta$) \cite{acta}.}
 \label{OO}
 \end{figure}
 temperature dependence of which defines structural transition. This is depicted in Fig.\ref{OO}. It should be clearly noted that the
 structural distortion is predominantly determined by the orbital ordering at X(Y) point; it is very weakly influenced
 by the orbital ordering at $\Gamma$ point which has very weak temperature dependence. This clearly shows that the orbital 
 ordering is very anisotropic. Now, question is why is that the orbital ordering at the (zone centre) $\Gamma$ point is so weak but finite and independent of temperature? This is also observed experimentally by Zhang et al., \cite{Zhang} and modelled as bond-order.
 We argue below this as manifestations of orbital anisotropy in presence of zone folding due to magnetic order. It is easy to
 envisage from the band structure in Fig.\ref{BS} that because of interband nesting E$_{d_{xz}}(k+Q_x)=-E_{d_{xz}}(k)$,  but 
 E$_{d_{yz}}(k+Q_x)=-E_{d_{xz}}(k)$ and E$_{d_{xz}}(k+Q_y)=-E_{d_{yz}}(k)$,  but  E$_{d_{yz}}(k+Q_y)=-E_{d_{yz}}(k)$. These would make the nematic order parameter (n$_{d_{xz}}$-n$_{d_{yz}}$) null if the nesting wave vectors Q$_x$, Q$_y$ were equivalent, but as it is not, it results in some small but finite quantity which is nearly independent of temperature. 
 This feature is indicative of the fact that the orbital ordering 'gap' would form a density wave 
and this along with the SDW state is interband in nature \cite{Ghoshepl}.
  \begin{figure}
    \centering
    \includegraphics [height=6.75cm,width=8.0cm]{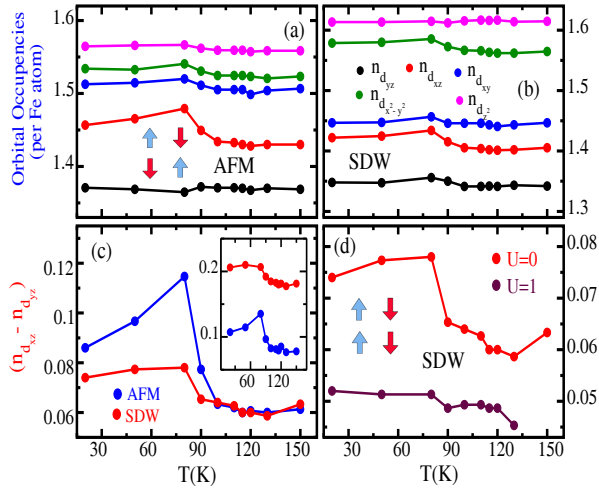}
    \caption{Total orbital occupancies per Fe atom (in real space) of various d orbitals of all Fe-atoms in the super-cell as a function of temperature for (a) AFM and 
    (b)
    SDW spin configuration indicated in the inset. (c) (n$_{d_{xz}}$-n$_{d_{yz}}$) and (n$_{d_{xz}}$-n$_{d_{yz}}$)+(n$_{d_{{x^2}-{y^2}}}$-n$_{d_{xy}}$) (inset) as a function of temperature
    for AFM (blue) and SDW (red) spin structures. (d) Thermal variation of 
  (n$_{d_{xz}}$-n$_{d_{yz}}$) for SDW spin configuration
    considering correlation (U=1) using GGA+U formalism and with out correlation.}
   \label{OOSC}
   \end{figure}
By now through above discussions it is nearly evident as to what is the source of 
temperature dependence of the orbital order parameter and may relate to the same of the nematic order parameter.
The Fe band energies presented in Figs. \ref{BS} may be written as, 
E(k)=$\sum_i  \epsilon_i (k)n_i (k); i=d_{xz} ,d_{yz},d_{xy},d_{x^2-y^2} ,d_{(3z^2-r^2)}$  and 
the corresponding eigenstates involving orbitals are $\Psi=\sum_ic_i\phi_i$. $\epsilon_i (k)$ 
and n$_i$(k)s are the band energies and occupation probabilities of 'i'th orbital $\phi_i$ respectively. 
The $\epsilon_i$(r)s being the Kohn-Sham orbit energies and the corresponding Fourier transformed $\epsilon_i$(k)s are independent of temperature (and magnetic interaction introduced later through I$_s$), whereas the orbital occupancies (or densities) n$_i$(k) are function of temperature. Therefore, lifting of degeneracy of the d$_{xz}$, d$_{yz}$ bands at the $\Gamma$ and X points, as the temperature is lowered below structural transition temperature is a consequence of the fact that their occupation probabilities become different ({\it i.e}, partial densities become unequal). Note, the energy gap between the  d$_{xz}$, d$_{yz}$ bands at the $\Gamma$ and X points is a function of temperature (see Fig. \ref{OO}). The temperature difference essentially originates from the temperature dependencies of n$_{d_{xz}}$, n$_{d_{yz}}$ and is proportional to n$_{d_{xz}}$-n$_{d_{yz}}$ (see Fig. \ref{OOSC}c). Therefore, it is desirable to calculate the temperature dependencies of the occupation probabilities of all the five d-orbitals of Fe from first principles calculation.
Such a rare calculation is presented in Fig. \ref{OOSC}.
 This quantity (n$_{d_{xz}}$-n$_{d_{yz}}$) represents inter orbital charge fluctuation or orbital fluctuation in short, 
 one of the important contender for nematic phase and is also called nematic order parameter.
Using super-cell of orthorhombic BaFe$_2$As$_2$ structure corresponding to $\sim$ 6 $\%$ Ru doping (cf. Fig.1) and two types of spin arrangements AFM and SDW (shown in the inset of Fig. \ref{OOSC}) first principles simulations of orbital occupancies are presented. 
 \begin{figure}
    \centering
    \includegraphics [height=3.8cm,width=8.0cm]{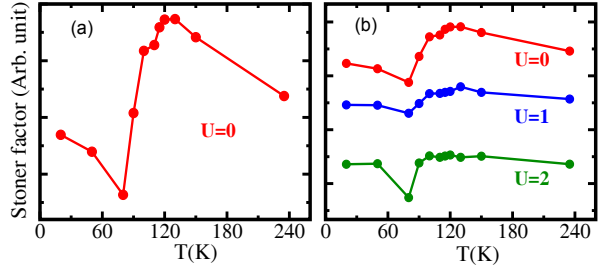}
    \caption{(a) Calculated Stoner factor ($I^{Fe}\times[N^{Fe}(E_F)]^2+I^{Ru}\times[N^{Ru}(E_F)]^2$) 
    as a function of temperature for 5$\%$ Ru doped BaFe$_2$As$_2$ systems and
    (b) thermal variation of Stoner
     factor for different U values as indicated in the figure.}
   \label{Stoner}
   \end{figure}
Why dope BaFe$_2$As$_2$ with Ru? Like hole doped 122 systems iso-electronic Ru doped 122 system 
also have inseparably same structural and magnetic transitions and the nematic phase in this system 
remain unexplored. This is unlike other iso-electronic P doping in place of As. 
It is particularly an interesting case, it is not clear as to where does 
the charge carrier go in case of Ru doping in place of Fe. Both the hole and 
electron Fermi pockets either remain unaltered or expands at an equal rate \cite{Dhaka,pla}).
 Furthermore, both the structural and magnetic transitions are 2nd order in nature 
 in case of underdoped Ru-122 system. Therefore, study of temperature dependence of orbital 
 fluctuation from first principles is of genuine interest.
In Fig.\ref{OOSC} orbital occupancies of d$_{xz}$ orbital (n$_{d_{xz}}$) 
modifies significantly with temperature compared to the other d orbitals specially d$_{yz}$ and d$_{xy}$ (but they also do
show substantial temperature dependence).
We have also calculated the difference in the occupancies of d$_{xz}$ and d$_{yz}$ orbitals {\it i.e},
 n$_{d_{xz}}$-n$_{d_{yz}}$ (nematicity) as a function of temperature for both AFM and 
 SDW spin configuration (see Fig. \ref{OOSC}c). Since, above structural transition there is no splitting between
 the d$_{xz}$ and d$_{yz}$ bands ({\it i.e,} $\epsilon_{xz}=\epsilon_{yz})$ temperature dependence of the nematic order
 parameter  n$_{d_{xz}}$-n$_{d_{yz}}$ is proportional to that of the orbital order. 
In other words, the nematic order essentially grow as orbital order which is responsible for orthorhombic 
transition. Therefore, this result may be interpreted as the first principles evidence 
of the fact that if orbital fluctuation is the primary order responsible for nematicity, then it is 
proportional to the orthorhombicity parameter \cite{Fernandes}. This is one of the 
inputs of all the theories involving Ginzburg Landau formalism. 
As already mentioned towards the end of theoretical method section that the method of calculation of occupation 
probabilities of U effected orbitals remains same as that of the ref. \cite{Cococcioni,Anisimov} which 
is also valid even in the limit U tending to zero. Therefore, occupation probabilities are obtained for a very small 
U$=0.01$eV (not exactly equal to zero) which makes up/down spin states different even in the tetragonal phase. 
As a result n$_{d_{xz}}$-n$_{d_{yz}}$ become very small but non-zero and has hardly any temperature dependence. 
Also, the difference in the nematic order parameter for SDW and AFM clearly indicates that it is a spin nematicity.
Furthermore, in the inset of Fig. \ref{OOSC}c, we depict the thermal variation of (n$_{d_{xz}}$-n$_{d_{yz}}$)+(n$_{d_{{x^2}-{y^2}}}$-n$_{d_{xy}}$) which also reproduces thermal behaviour of orthorhombicity corresponding to orbital fluctuation involving all four d-orbitals. Thus, in contrast to the usual
  belief in literature, nematic order parameter which is defined as n$_{d_{xz}}$-n$_{d_{yz}}$, perhaps involve all 
  other $d $-orbitals as well. 
 \begin{figure}
   \centering
   \includegraphics [height=8.8cm,width=8.0cm]{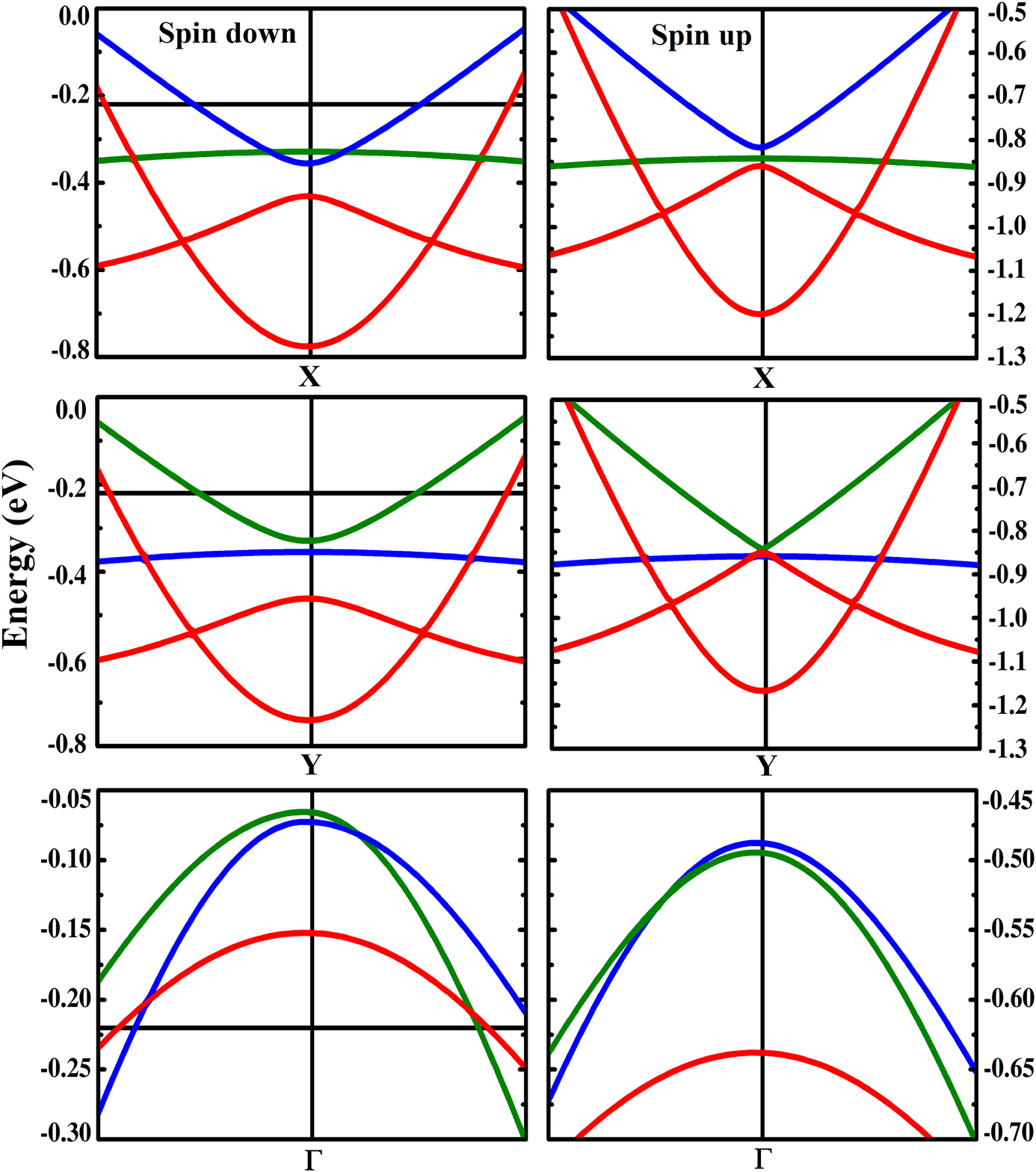}
   \caption{Calculated band structure of 5$\%$ Ru doped BaFe$_2$As$_2$ systems at 20 K
   with I$_s$ = 1 for down spin (left) and up spin (right) electrons around X, Y and $\Gamma$ points.
   We assign red green and blue colours to d$_{xy}$, d$_{yz}$ and d$_{xz}$ orbitals respectively.}
  \label{BSFSM}
  \end{figure}
The nematic order parameters show substantial modifications near the temperature where magnetic fluctuation is very strong (see Fig.\ref{OOFSM}). This provides a distinct evidence to the fact that probably magnetic fluctuation, orbital fluctuation, the nematicity  are interdependent (we discuss this issue below).
 Also it should be noted that the orbital occupancies of d$_{xz}$ orbital
 is always greater than that of the d$_{yz}$ orbital {\it i.e.,} n$_{d_{xz}}$$>$n$_{d_{yz}}$.
 This propound ferro-orbital ordering. Partial densities n$_{xz}$, n$_{yz}$ being unequal would correspond to different bonding along 
 $x$ and $y$ directions, a mark of nematicity due to orbital anisotropy. 
 Unlike cuprates it is experimentally well established that 122 family of Fe-based SCs
 are weakly correlated. In Fig. \ref{OOSC}(d) temperature dependence of n$_{d_{xz}}$-n$_{d_{yz}}$
 has been shown with small on site correlation (U=1) and with out correlation. It is clear that correlation reduces
 orbital order and so the magnetic fluctuation (see Fig. \ref{Stoner}b).
Magnetic fluctuations play an important role in these family of Fe-based SCs. Stoner factor is the measure
of these magnetic fluctuation. Stoner factor of this compound can be defined as
 $I^{Fe}\times[N^{Fe}(E_F)]^2+I^{Ru}\times[N^{Ru}(E_F)]^2$, where $N^{Fe}(E_F)$ and $N^{Ru}(E_F)$ are the density of states
 at the Fermi level from Fe and Ru atoms respectively \cite{Stoner,Wang}. 
 The value of Stoner parameters $I^{Fe}$ and $I^{Ru}$ are taken 
 from ref \cite{Stoner,Yan}. We have calculated the Stoner factor 
 as a function of temperature and displayed in Fig.\ref{Stoner}a.
 As temperature changes, partial density of states of Fe and 
 Ru at the Fermi level (E$_F$) get modified due to substantial moderation 
 of Fe-As hybridization. This is the root cause of temperature dependence of Stoner factor.
 This observation is very much consistent with recent experimental findings \cite{matan}.
 In Fig. \ref{Stoner}b thermal variation of Stoner factor has been depicted with different values of U (strength of electron correlation). It is clear that with increasing electron repulsion (system would 
prefer stable antiferromagnetic configuration) magnetic fluctuation is actually decreasing 
(so is the nematic order) which enhances further our suspect that magnetic 
 fluctuation triggers orbital fluctuation and nematic phase.\\

For this purpose, we tune magnetic interaction manually by introducing integrated spin density parameter as 
explained 
above and see its influence on electronic structure. 
While keeping initial AFM spin structure we provide constraint through the integrated spin density parameter for 
spin polarized calculations which induces some magnetic moment in the system. This method is a simple extension 
of standard LSDA formalism where total energy as a function of moment can be obtained. After complete spin 
polarized calculation being performed, the main idea is to perform a single point energy calculation keeping 
the total moment constrained to some fixed small but finite non zero values. This is equivalent to generating 
a weak Zeeman field (see also the form of the integrated spin density parameter). In this way by keeping the AFM 
spin configuration, magnetic interaction can be introduced to the system by simply fixing the total (difference 
in up and down spin) magnetic moment of the system through integrated spin density parameter.
Actually, from electronic structure calculation 
we calculate orbital ordering again in presence of magnetic interaction (see below) and see that orbital anisotropy is enhanced, and hence the nematicity.
\begin{figure}
 \centering
 \includegraphics [height=4cm,width=8.5cm]{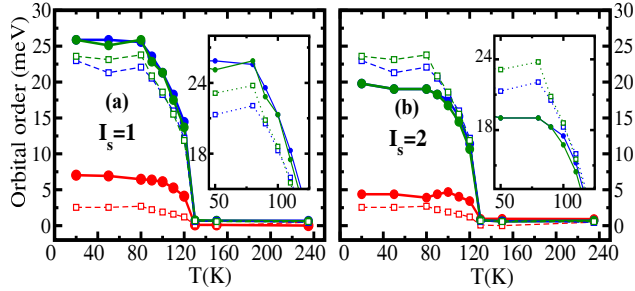}
 \caption{Calculated orbital order (in meV) around X (blue), Y (green) and $\Gamma$ (red) 
 points as a function of temperature for (a) I$_s$ = 1 
 and (b) I$_s$ = 2 for 5$\%$ Ru doped BaFe$_2$As$_2$ systems. Hollow symbols represent data from Fig. \ref{OO}.}
\label{OOFSM}
\end{figure}
In presence of finite integrated spin density, the band of up spin electrons and band of down spin electrons split.
 We observed that for I$_s$ = 1 and 2, one of the bands (up spin) goes deep below the Fermi level. 
 So only one of the spin electronic bands are
 contributing significantly at the Fermi level. In Fig. \ref{BSFSM} we have shown the band structures of
 5$\%$ Ru doped BaFe$_2$As$_2$ systems at 20 K, I$_s$ = 1 for down spin (left) and up spin (right)
 electrons around different k points (X, Y and $\Gamma$ points). Notably, around X, Y and $\Gamma$ points all 
the orbitals are ordered differently in contrast to that in Fig. \ref{BS}.
Around X as well as around $\Gamma$ point the energy ordering of d$_{xz}$ and d$_{yz}$ orbitals
are exactly opposite to each other for up spin and down spin bands. This leads to 
spin-polarized orbital orderings (possibly orbital density wave).
We do all the same exercises (in all the figures above) for K doped as well as P doped 122 systems 
and found that the orbital ordering is a common phenomena (there are differences in details and will 
be reported elsewhere).
In Fig. \ref{OOFSM} we have presented temperature variation of orbital order
around X, Y and $\Gamma$ points for I$_s$ = 1 (Fig. \ref{OOFSM}a) and I$_s$ = 2 respectively 
(Fig. \ref{OOFSM}b) after extracting the required information from Fig. \ref{BSFSM}.
In case of I$_s$ = 1 the orbital order around $\Gamma$ increases to about 3 fold compared
to the case where we optimized the total spin of the system (represented by hollow symbols). 
However, the orbital ordering around the zone corners X(Y)
are less affected. When the integrated spin density is increased to 1 from 0, it stabilizes the 
underlying SDW, but when I$_s$ is further increased to 2, the underlying SDW ordering will be 
strongly suppressed due to ferromagnetic nature of the I$_s$. Hence, orbital ordering is 
strongly coupled to the underlying magnetic fluctuation (it enhances orbital fluctuation or nematic order parameter) 
and our study thus is complementary to recent studies 
\cite{Fernandes,Yi,Zhang,Luo1,Luo2} on nematic phase.

\section{Conclusion}

We establish the microscopic relationship between the orbital order, structural transition and 
nematic order in 122 family of Fe-based superconductors. While electronic orbital anisotropy gives 
rise to orbital order, temperature dependence of the orbital order is found to be exactly same as 
that of the orthorhombicity, indicating orbital ordering is responsible for structural transition. 
Temperature dependence of the orbital order is proportional to the temperature dependence of the 
nematic order (n$_{d_{xz}}$-n$_{d_{yz}}$). This indicates that the nematic order 
grows as orbital order in the orthorhombic phase. We have explicitly evaluated the temperature 
dependencies of orbital occupancies of all Fe-d-orbitals. Almost all the d-orbitals show substantial
charge fluctuations in the orthrhombic phase, indicating that the actual definition of nematic order 
parameter may be more complicated. The nematic order parameter is found to show temperature dependence 
close to the onset of magnetic fluctuation, obtained rigorously through evaluation of 
Stoner factor. When magnetic fluctuations are enhanced, the orbital fluctuations are also enhanced 
and vice versa establishing their couplings. Spin-polarised orbital ordering revealed from this work 
would be experimentally observable. Finally, our work supports magnetic origin of nematicity in 122 
family of Fe-based superconductors. We believe this work would generate further interest in theoretical 
as well as experimental studies.\\ 

{\bf Acknowledgements} \\

We thank Dr. A. Bharathi and Dr. A. K. Sinha for discussion on experimental 
aspects. We thank Dr. P. A. Naik and Dr. P. D. Gupta for their encouragement in this work. One of 
us (SS) acknowledges the HBNI, RRCAT for financial support and encouragements.\\ 

{\bf Additional information} \\

{Competing financial interests:} The authors declare no competing financial interests. \\





\end{document}